\documentclass[aps,prl,onecolumn,amsmath,amssymb,superscriptaddress]{revtex4}
\usepackage{graphicx}
\usepackage{amssymb}
\usepackage{natbib}
\usepackage{color}
\usepackage{soul}

\newcommand{\beq}{\begin{eqnarray}}
\newcommand{\eeq}{\end{eqnarray}}

\begin{document}

\title{Spin excitations and the Fermi surface of superconducting FeS}

\author{Haoran Man}
\affiliation{Department of Physics and Astronomy
and Center for Quantum Materials, 
Rice University, Houston, Texas 77005, USA}

\author{Jiangang Guo}
\affiliation{Department of Physics and Astronomy
and Center for Quantum Materials, Rice University, Houston, Texas 77005, USA}

\author{Rui Zhang}
\affiliation{Department of Physics and Astronomy
and Center for Quantum Materials, Rice University, Houston, Texas 77005, USA}

\author{Rico U. Sch\"{o}nemann}
\affiliation{National High Magnetic Field Laboratory, Tallahassee, Florida 32310, USA}

\author{Zhiping Yin}
\email{yinzhiping@bnu.edu.cn}
\affiliation{Center for Advanced Quantum Studies and Department of Physics, Beijing Normal University, Beijing 100875, China}

\author{Mingxuan Fu}
\affiliation{Department of Physics and Astronomy, Johns Hopkins University, Baltimore, MD 21218, USA}

\author{M. B. Stone}
\affiliation{Quantum Condensed Matter Division, Oak Ridge National Laboratory, Oak Ridge, Tennessee 37831, USA}

\author{Qingzhen Huang}
\affiliation{NIST Center for Neutron Research, National Institute of Standards and Technology, Gaithersburg, Maryland 20899, USA}

\author{Yu Song}
\affiliation{Department of Physics and Astronomy
and Center for Quantum Materials, Rice University, Houston, Texas 77005, USA}

\author{Weiyi Wang}
\affiliation{Department of Physics and Astronomy
and Center for Quantum Materials, Rice University, Houston, Texas 77005, USA}

\author{David Singh}
\affiliation{Department of Physics and Astronomy, University of Missouri, Columbia, MO, USA}

\author{Felix Lochner}
\affiliation{\textsuperscript{}Institut f\"ur Theoretische Physik III, Ruhr-University
	Bochum, D-44801 Bochum, Germany}
	
\author{Tillman Hickel}
\affiliation{\textsuperscript{}Max-Planck Institut f\"ur Eisenforschung, D\"usseldorf, Germany}

\author{Ilya Eremin}
\affiliation{\textsuperscript{}Institut f\"ur Theoretische Physik III, Ruhr-University
	Bochum, D-44801 Bochum, Germany}
	
\author{Leland Harriger}
\affiliation{NIST Center for Neutron Research, National Institute of Standards and Technology, Gaithersburg, Maryland 20899, USA}

\author{Jeffrey W. Lynn}
\affiliation{NIST Center for Neutron Research, National Institute of Standards and Technology, Gaithersburg, Maryland 20899, USA}

\author{Collin Broholm}
\affiliation{Department of Physics and Astronomy, Johns Hopkins University, Baltimore, MD 21218, USA}
\affiliation{NIST Center for Neutron Research, National Institute of Standards and Technology, Gaithersburg, Maryland 20899, USA}

\author{Luis Balicas}
\email{balicas@magnet.fsu.edu}
\affiliation{National High Magnetic Field Laboratory, Tallahassee, Florida 32310, USA}

\author{Qimiao Si}
\affiliation{Department of Physics and Astronomy
and Center for Quantum Materials, Rice University, Houston, Texas 77005, USA}

\author{Pengcheng Dai}
\email{pdai@rice.edu}
\affiliation{Department of Physics and Astronomy
and Center for Quantum Materials, Rice University, Houston, Texas 77005, USA}
\affiliation{Center for Advanced Quantum Studies and Department of Physics, Beijing Normal University, Beijing 100875, China}

\maketitle
\textit{Correspondence and
requests for materials should be addressed to 
Z.P.Y (yinzhiping@bnu.edu.cn), L.B. (balicas@magnet.fsu.edu), or 
P.D. (e-mail: pdai@rice.edu)}

\section{Abstract}
High-temperature superconductivity occurs near antiferromagnetic instabilities and nematic state. Debate remains on the origin of nematic order in FeSe and its relation with superconductivity. Here, we use transport, neutron scattering and Fermi surface measurements to demonstrate that hydro-thermo grown superconducting FeS, an isostructure of FeSe,  is a tetragonal paramagnet without nematic order and with a quasiparticle mass significantly reduced from that of FeSe. Only stripe-type spin excitation is observed up to 100 meV. No direct coupling between spin excitation and superconductivity in FeS is found, suggesting that FeS is less correlated and the nematic order in FeSe is due to competing checkerboard and stripe spin fluctuations.

\section{Introduction}

High-transition temperature (high-$T_c$) superconductivity in copper oxides and iron-based materials occurs near checkerboard and stripe antiferromagnetic (AF) instabilities, respectively \cite{scalapino,dai,si2016}. Although there is also ample evidence for the existence of a nematic order,  where a translationally invariant metallic phase
spontaneously breaks rotational symmetry 
\cite{fradkin,fisher,Fernandes2011,anna,xylu14}, and nematic quantum critical point (QCP) near optimal superconductivity in iron-based superconductors \cite{Dai2009,Kuo2016}, much is unclear concerning its microscopic origin and relationship to superconductivity \cite{dai,si2016}. In particular, recent debates focus on whether the nematic order in superconducting FeSe below the tetragonal-to-orthorhombic transition temperature $T_s=91$ K without static AF order \cite{McQueen,baek15,anna15} is due to competing magnetic instabilities or to orbital ordering \cite{Rahn,WangQ,WangQa,YuR,WangF,Glasbrenner,CaoHY,Chubukov15,Yamakawa}.
Here, we use transport, neutron scattering and Fermi surface measurements to demonstrate that
superconducting FeS, an isostructure of FeSe \cite{XLai,borg},  is a tetragonal paramagnet without nematic order 
and with a quasiparticle mass significantly 
reduced from 
that of FeSe.
Our neutron scattering experiments in the energy regime below 100 meV reveal only 
stripe-type spin fluctuations in FeS that are not directly coupled to superconductivity.
These properties suggest
that FeS is a weakly correlated analog of FeSe and, moreover, that the
nematic order in FeSe is 
due to the frustrated magnetic interactions underlying the
competing checkerboard and stripe spin fluctuations \cite{WangQa,YuR,WangF}.

A key to understanding the physics of the iron-based superconductors is to determine the role of magnetism and electronic nematic phase to superconductivity \cite{scalapino,dai,si2016,fisher,Fernandes2011,anna}.
In a typical AF ordered iron-pnictide, a tetragonal-to-orthorhombic lattice distortion $T_s$ occurs at temperatures above or at the AF ordering temperature $T_N$ \cite{dai}, and the nematic phase is observed in the
paramagnetic orthorhombic phase between $T_s$ and $T_N$ \cite{fisher,Fernandes2011,anna}.
Although iron chalcogenide FeSe single crystals [Fig. 1(a) and 1(b)] also undergo
a nematic transition at $T_s$ and become superconducting at $T_c=9.3$ K \cite{McQueen}, the low-temperature static AF ordered phase is absent \cite{baek15,anna15}. This has fueled debates concerning the role of AF order and spin fluctuations to the nematic phase and superconductivity \cite{baek15,anna15,Rahn,WangQ,WangQa,YuR,WangF,Glasbrenner,CaoHY,Chubukov15,Yamakawa}. Initially, nuclear magnetic resonance (NMR) experiments on FeSe suggested that magnetism plays no role in its nematic
transition \cite{baek15,anna15}.  However, subsequent neutron scattering measurements reveal strong low-energy spin fluctuations at the stripe AF ordering wave vector and a resonance coupled to superconductivity \cite{Rahn,WangQ}, similar to spin fluctuations in the iron pnictides \cite{dai}.  In addition, recent 
spin excitation measurements suggest that the nematic transition in FeSe is due to a competition between the checkerboard and the stripe spin 
fluctuations at AF wave vectors $(1,1)$ and $(1,0)$, respectively [Fig. 1(c) and 1(d)]  \cite{WangQa}, consistent 
with the frustrating magnetic interactions \cite{YuR,WangF}. 
In this picture, one would expect that S-substituted FeSe$_{1-x}$S$_x$, which reduces $T_s$ and lattice orthorhombicity
\cite{Watson15,LRWang}, should have reduced spin fluctuations associated with the checkerboard order.  
As FeS single crystals are
isostructural to FeSe but
with a reduced $T_c=4.3$ K, it 
should allow a direct comparison with FeSe \cite{Rahn,WangQ,WangQa}, and thus elucidate
the role of spin fluctuations to the nematic phase and to superconductivity.

\section{Results}

Here, we use
 transport (Fig. 1),
neutron scattering (Figs. 2 and 3), quantum oscillation experiments (Fig. 4), as well as 
density function theory (DFT) \cite{Subedi08} and DFT combined with dynamical mean field theory (DMFT) calculations \cite{Yin,Yin14} to study single crystals of FeS \cite{SI}. 
To search for the presence of a nematic phase in FeS, we performed elastoresistance measurements on single crystals of FeS and BaFe$_{1.97}$Ni$_{0.03}$As$_2$ \cite{xylu16} using 
a piezo electric device [Fig. 1(g)] \cite{Kuo2016}. 

Figure 1(h) compares thestrain dependence of the elastoresistance at different temperatures for 
FeS, FeSe, and BaFe$_{1.97}$Ni$_{0.03}$As$_2$, respectively. While there is a clear resistivity anisotropy for FeSe and BaFe$_{1.97}$Ni$_{0.03}$As$_2$, indicative of a nematic phase, FeS reveals no anisotropy in measurements of the elastoresistance from 5 K to 105 K.
We therefore conclude that FeS has no nematic order, which is consistent with the previous reports on FeS
\cite{XLai,borg} and  
with the notion that the nematic phase vanishes 
for FeSe$_{1-x}$S$_x$ for $x\geq 0.17$ \cite{Hosoi,XX1,XX2}.
The results from the transport measurements are complemented by those from elastic neutron scattering measurements, 
which reveal that the system is paramagnetic at all temperatures \cite{SI}, suggesting that the previous observation of magnetic order in FeS is likely due 
to impurity phases \cite{Holenstein,Kirschner}.

Having established the absence of any nematic order in FeS, we turn to probing the 
spin excitation spectrum by inelastic neutron scattering experiments.
Figure 2 summarizes our neutron time-of-flight measurements on FeS to determine the overall wave vector and energy dependence of the spin fluctuations \cite{SI}. 
For these measurements, we use orthorhombic unit cell notation and define momentum transfer ${\bf Q}$ in three-dimensional (3D) reciprocal space in \AA$^{-1}$ as $\textbf{Q}=H\textbf{a}^\ast+K\textbf{b}^\ast+L\textbf{c}^\ast$, where $H$, $K$, and $L$ are Miller indices and
${\bf a}^\ast=\hat{{\bf a}}2\pi/a$, ${\bf b}^\ast=\hat{{\bf b}}2\pi/b$, ${\bf c}^\ast=\hat{{\bf c}}2\pi/c$
[Fig. 1(c) and 1(d)].
Our single crystals are aligned with the $c$-axis along the incident beam and with the $a$-axis in the horizontal plane.  In this geometry, we expect that the checkerboard and stripe AF correlations occurs at $(\pm 1,\pm1)$ and $(\pm1,0)$ in-plane wave vectors, respectively.
Figure 2(a)-2(d) shows the spin excitations of FeS at energy transfers of $E=20\pm 4, 40\pm 5, 50\pm 7,$ and $59\pm 7$ meV, respectively.  
In all cases, we see transversely elongated spin excitations centered around the stripe wave vector $(1,0)$ with no obvious magnetic signal 
at the checkerboard wave vector $(1,1)$.  Since magnetic scattering is normalized to absolute units using a vanadium standard \cite{dai}, we can quantitatively compare the results with those of FeSe \cite{WangQ,WangQa}.
Figures 2(e)-2(h) show the transverse cuts for FeS (solid circles) and FeSe (solid lines) corresponding to energies in Figs. 2(a)-2(d)
along the $[1,k]$ direction [see red dashed lines in Fig. 2(a) for scan direction].  
Integrating the scattering over the same energy interval, we see that the FeS scattering is much weaker, and we do not observe 
magnetic scattering associated with the checkerboard correlations for energies below 100 meV, 
 in contrast with the clear magnetic scattering of FeSe at $(1,1)$ as marked by vertical arrows in Fig. 2(e)-2(h).  
Figure 2(i) compares the energy dependence of the local dynamic susceptibility $\chi^{\prime\prime}(E)$, defined as the dynamic susceptibility integrated over the dashed white box in
Fig. 2(a) \cite{dai}, for both FeS and FeSe \cite{WangQa}.  
Within the energy region probed, $\chi^{\prime\prime}(E)$ increases with increasing energy but has about a quarter of the intensity of FeSe [Fig. 2(i)].

To determine if spin excitations in FeS couple to superconductivity, we carried out temperature dependence measurements of the low-energy spin fluctuations near the stripe ordering wave vector $(1,0)$.  For this purpose, single crystals of FeS were aligned in the $[H,0,L]$ scattering plane, and maps of scattering intensity at different energies above and below $T_c$ were measured using a cold neutron spectrometer. Figures 3(a)-3(d) show background subtracted scattering maps at $E=0.75$, 2, 4 and 6 meV, respectively, well below $T_c$ at $T=1.5$ K.  In all cases, we see rod-like scattering centered at $(1,0,L)$ with extended scattering along the $L$ direction, consistent with short-range $c$-axis spin correlations.  In the case of FeSe, a neutron spin resonance coupled to superconductivity was found near $E_r=4$ meV, which correspond to approximately $5.3k_BT_c$ where $k_B$ is the Boltzmann constant, at $(1,0)$ \cite{Rahn,WangQ}.  Since the $T_c$ of FeS is about half of that of FeSe, the resonance in FeS should be present around $E_r\approx 2$ meV.  To accurately determine the temperature dependence of the dynamic susceptibility near $(1,0)$, we integrate the scattering around $(1,0)$ along the $L$ direction, and then fit the profile to a Gaussian on a linear background [see inset in Fig. 3(e)].  After correcting for the Bose factor, we show in Fig. 3(e) the temperature dependence of the dynamic susceptibility $\chi^{\prime\prime}(E)$ near the wave vector $(1,0)$.  The energy dependence of $\chi^{\prime\prime}(E)$ is weakly temperature dependent below about 10 meV and shows no evidence for a neutron spin resonance expected around $E_r\approx 2$ meV.

The contrast in the spin dynamics between FeS and FeSe is striking and provides the clue to the physics of both systems. 
We start from the observation that,
as in the case of P-for-As substitution \cite{Dai2009}, 
the reduction of Fe-pnictogen distance on moving from FeSe to FeS facilitates electron hopping, 
and thus reduces the electron correlations [Fig. 1(b)],  as seen in spin excitations of BaFe$_2$(As$_{0.7}$P$_{0.3}$)$_2$ \cite{xx3}. 
The notion that FeS is a less correlated analogue of FeSe is qualitatively consistent with our conclusion
 that the spin spectral weight at low energies is much reduced in FeS compared to FeSe
 [Fig. 2(i)]. 
 
The stoichiometric nature of FeS facilitates both quantum oscillation measurements and electronic structure calculations, thereby providing the opportunity to address 
  the correlation physics in a more quantitative way. We therefore 
  turn to the understanding of both the Fermi surface and the effective quasiparticle mass. Figure 1(e) shows the calculated Fermi surfaces of FeS using combined DFT and DMFT \cite{SI}.  Comparing with schematics of the measured Fermi surfaces of FeSe in Fig. 1(f) \cite{Watson15}, substituting S for Se in FeSe induces the $d_{xy}$ orbital hole pocket near $(1,1)$ and changes the properties of the hole pockets near the $\Gamma$ point $(0,0)$ [Fig. 1(e)].
 To quantitatively
determine the differences in the Fermi surfaces of FeS and FeSe, we performed torque magnetometry and resistivity
measurements under high magnetic fields. 
Figure 4 summarizes the quantum oscillatory phenomena observed on FeS investigated through torque magnetometry and resistivity measurements under fields as high as $\mu_0H = 35\,\mathrm{T}$ in resistive Bitter magnets equipped with either a $^{3}\mathrm{He}$ refrigerator or $^{4}\mathrm{He}$ cryostat. Resistivity measurements were performed on a sample characterized by a residual resistivity ratio ($RRR = R_{300\,\mathrm{K}}/R_{6\,\mathrm{K}}$) of 41, using a standard four wire technique, while torque was measured through a cantilever beam set-up whose deflection was determined capacitively \cite{SI}. We were able to observe well pronounced Shubnikov-de Haas (SdH) and de Haas-van Alphen (dHvA) oscillations in the resistance and in torque measurements, respectively. Typical dHvA and SdH oscillations and their respective Fast Fourier Transformations (FFT's) for $H\parallel c-$axis are shown in Figs. 4(a) and 4(b), respectively. Although their amplitudes differ, most of the SdH frequencies observed below $1\,\mathrm{kT}$, which are indicated by the peaks labeled as $\alpha$, $\beta$, $\kappa$, $\delta$ and $\epsilon$, are reproduced in the dHvA spectrum. Only $\nu$ and $\gamma$ are not visible in the dHvA data. Furthermore, the prominent dHvA peaks at $F = 370\,\mathrm{T}$ and $400\,\mathrm{T}$ seem to be suppressed in the SdH data, which is attributable to the lower temperature for the torque measurements. Here, it is important to emphasize that the SdH-effect is superimposed onto an electrical transport quantity (resistivity) which is driven by scattering processes, while the dHvA one is superimposed onto a thermodynamic variable (magnetic susceptibility) which, in a metal is dependent upon the density of states at the Fermi level. Therefore, it is not surprising that the relative amplitude between peaks observed in the FFT spectra is technique dependent. In addition, different crystals from a given synthesis batch are likely to display variations in mobility. This should affect the detection of some of the orbits and hence also produce comparative differences in the FFT spectra collected from the different crystals, as seen in our experiments.

 The effective mass $\mu$ of the different orbits can be extracted from the temperature dependence of the FFT amplitude as depicted in Fig. 4(c). The decrease of the FFT amplitude with increasing temperature is described by the Lifshitz-Kosevich damping factor $R_{T} = \pi\lambda/\sinh(\pi\lambda)$. Considering only the first harmonic, one gets $\lambda = 2\pi k_{\mathrm{B}}T/\beta H$, where $\beta \propto 1/\mu$. This analysis yields effective masses of $1.1(1)m_{0}$ for the $\alpha$, $\beta$ and $\kappa$ orbits as well as $1.7(1)$, $1.8(2)$, $1.9(2)$ and $1.8(2)m_{0}$ for the $\delta$, $\epsilon$, $\nu$ and $\gamma$ orbits. Thus charge carriers in FeS have lower effective masses than those of FeSe whose masses range from 1.9 to $7.2m_{0}$ \cite{terashima_anomalous_2014}.Notice that we obtain somewhat heavier masses for the $\alpha$ and $\beta$ orbits than the values reported in Ref. \onlinecite{Terashima_FeS}. We re-analyzed our data by, for instance, extracting the effective masses from different field windows. However, we found that this does not explain the difference between the effective mass values extracted from both studies.

This is consistent with our DFT+DMFT calculations with mass enhancement $m^\ast/m_{band}$ of 1.9/1.6 for $t_{2g}$/$e_g$ orbitals in FeS, 
which is much smaller than that in FeSe \cite{Yin}.
The whole angular dependence of the SdH and dHvA frequencies as a function of $\theta$ is shown in Fig. 4(d), where $\theta$ denotes the angle between $H$ and the crystallographic $c$-axis. Based on the dHvA measurements, we observe a multitude of frequencies especially in the region between $0.3$ and $0.6\,\mathrm{kT}$ as well as at least three additional Fermi surface pockets with $F\geq 1\,\mathrm{kT}$. While tracking the individual frequencies that belong to certain Fermi surface sheets is a difficult task in the dHvA data, the picture seems to become clearer for the SdH oscillations. Nevertheless, we were not able to observe SdH oscillations for $\theta > 30^{\circ}$. The lines depicted in Fig. 4(d) are intended to provide a hint on the evolution of the frequencies as a function of $\theta$. However, a precise comparison with band structure calculations is required to associate the observed frequencies with specific Fermi surface sheets \cite{SI}. Band structure calculations find that the Fermi surface consists of two-dimensional (2D) cylindrical Fermi surface sheets at the center and at the corners of the Brillouin zone, respectively \cite{Subedi08}. 2D orbits would lead to a $F \propto (\cos(\theta))^{-1}$ dependence which are not clearly observed here. Although the angular dependence of some of the frequencies (e. g. $\alpha$ and $\beta$) could match a cylindrical Fermi surface, the bulk of the observed frequencies are clearly 3D in character and cannot be described by the currently available band structure calculations. A recent report on the SdH on FeS crystals detected only the two main peaks observed in our FFT spectra, probably because the measurements were performed at much lower fields \cite{Terashima_FeS}. However, the authors conclude that the Fermi surface of FeS has a 2D character in contrast to our observations.
Nevertheless, in their study the SdH oscillations were observed in a quite narrow angular range, i.e. $\Delta \theta \sim \pm 30^{\circ}$ with respect to the \emph{c}-axis, which is not a wide enough range to reach a definitive conclusion on the dimensionality of its Fermi surface. On the other hand, the observation of two of the same frequencies, or cross sectional areas, in samples grown by different groups further confirms that we are detecting the intrinsic Fermi surface of FeS.

\section{Discussion}

In an attempt to further understand the observed quantum oscillations in Figs. 4(a)-4(d), we carried out first-principles DFT plus single-site DMFT calculations in the paramagnetic phase of FeS, using the experimentally determined FeS crystal structure \cite{SI} and Hubbard $U=5.0$ eV and Hund's $J=0.8$ eV. 
When computing the 3D Fermi surface and the dHvA frequencies, we further incorporated the corrections from the long-range exchange interaction by shifting the 
hole (electron) Fermi surface down (up) by 50 meV. The calculated 3D Fermi surfaces are shown in Fig. 4(e). In particular, the middle hole Fermi surface and the two electron Fermi surfaces are quite 3D like, with large variation of the pocket size (cross section along the $[0,0,1]$ direction) along the $k_z$ direction. As shown in Fig. 4(f), the DFT+DMFT calculated dHvA frequencies agree well with experimental values.
We further assign each dHvA frequency to its corresponding position on the 3D Fermi surface \cite{SI}.

The reduced strength of the electrons correlations in FeS compared to FeSe also provides the understanding of the contrast in the spin dynamics of FeS to those of FeSe. Figures 4(g) and 4(h) show the energy dependence of the ground state magnetic scattering $S({\bf Q},E)$ for FeS and FeSe, respectively, calculated through a combination of DFT and DMFT methods \cite{Yin,Yin14}.  
The main conclusion from these calculations is that the spin excitations are much more energetic for FeS than for FeSe, with the 
strongest scattering for FeSe occurring below ~170 meV, while for FeS they extend to well beyond 400 meV similar 
to the case of iron phosphites  \cite{Yin14}. 

It is also instructive to compare the spin dynamics of the superconducting state in FeS 
with the results on FeSe and iron pnictide superconductors.  
For most iron-based superconductors, the appearance of superconductivity is coupled with changes in the spin excitations with the opening of a spin gap, and inducing a neutron spin resonance near the stripe AF wave vector \cite{dai}.  The presence of a resonance has mostly been interpreted as due to quasiparticle excitations between the hole-Fermi surfaces near the $\Gamma-$point and the electron Fermi surfaces near $(1,0)$ as a consequence of Fermi surface nesting \cite{dai}. Given the hole and electron Fermi surfaces in FeS [Fig. 1(e)] and FeSe  [Fig. 1(f)], one would expect the presence of spin fluctuations in both materials at
the commensurate stripe AF wave vector $(1,0)$.  
Our finding that FeS is
a weakly correlated analog of FeSe
provides a natural understanding of the lack of a neutron resonance. 
More quantitatively, from magnetic and transport measurements, it was argued that FeSe is deep inside Bardeen-Cooper-Schrieffer (BCS)
 and
Bose-Einstein-condensate (BCS-BES) cross-over regime, where the ratio of superconducting gap $\Delta$ to Fermi energy $\epsilon_F$ is of the order of unity \cite{Kasahara,Kasahara16}. 
From the experimentally obtained values for the SdH frequencies $F$ and the effective masses $\mu$ in FeS,
we can estimate the Fermi energy $\epsilon_F$ by using: $\epsilon_F = \hbar^{2}k_{\mathrm{F}}^{2}/2\mu$, $A = \pi k_{\mathrm{F}}{^2}$ and $F = \hbar A/2\pi e$.  Assuming that the
superconducting gap $\Delta$ can be estimated by using the BCS formula for a weakly coupled superconductor for FeS: $\Delta(T\rightarrow 0) = 1.764\,k_{\mathrm{B}}T_{\mathrm{c}} = 0.65\,\mathrm{meV}$ with $T_{\mathrm{c}}=4.3\,\mathrm{K}$, we can calculate the ratio of superconducting gap to Fermi energy as shown in the table below. 
It clearly shows
that the electron pairing in FeS is much closer to a 
BCS superconductor, again in line with our finding of a correlation strength in FeS that is considerably reduced than that of FeSe.

\vspace{1cm}
\begingroup
\setlength{\tabcolsep}{10pt} 
\renewcommand{\arraystretch}{1.5} 
\begin{table}
\begin{center}
	\begin{tabular}{l l l l l l l}
	\hline
		Branch & $F(\mathrm{kT})$ & $\mu/\mu_{0}$ & $A(\% \mathrm{BZ})$ & $k_{\mathrm{F}}(\mathrm{\AA^{-1}})$ & 	$\epsilon_F(\mathrm{meV})$ & $\Delta(T\rightarrow 0)/\epsilon_F$\\
		\hline
		$\alpha$ & 0.15 & 1.1 & 0.49 & 0.068 & 15.8 & 0.041\\
		$\beta$ & 0.21 & 1.1 & 0.67 & 0.079 & 21.7 & 0.029\\
		$\kappa$ & 0.29 & 1.1 & 0.96 & 0.094 & 30.9 & 0.021\\
		$\delta$ & 0.46 & 1.7 & 1.48 & 0.12 & 31.1 & 0.021\\
		$\epsilon$ & 1.07 & 1.8 & 3.48 & 0.18 & 68.8 & 0.0094\\
		$\nu$ & 1.40 & 1.9 & 4.55 & 0.21 & 85.1 & 0.0076\\
		$\gamma$ & 1.89 & 1.8 & 6.14 & 0.24 & 121 & 0.0054\\
		$2 \delta$ & 0.92 & $2.9 \pm 0.4$ & 2.96 & 0.17 & 38 & 0.017\\
		\hline
	\end{tabular}
\caption{\textbf{Summary of experimental data extracted from both the de-Haas-van-Alphen and the Shubnikov-de-Haas-effect}. Here $\alpha$, $\beta$, ... etc. stands for the frequencies observed in the FFT spectra, $\mu$ stands for the effective mass in units of the free electron mass $\mu_{0}$, $A(\% \mathrm{BZ})$ for the area of the cyclotronic orbit relative to the area of the Brillouin zone, $k_{\mathrm{F}}(\mathrm{\AA^{-1}})$  for the corresponding Fermi vector, $\epsilon_F(\mathrm{meV})$ corresponds to the associated Fermi energy and $\Delta(T\rightarrow 0)/\epsilon_F$ for the resulting ratio of the superconducting gap to the Fermi energy.}
\end{center}
\end{table}
\endgroup

To summarize, our inelastic neutron scattering experiments below 100 meV indicate that the spin excitations in FeS occur at the stripe AF wave vector $(1,0)$ with no observable signal at the checkerboard ordering wave vector $(1,1)$, and are much weaker than those of FeSe (Fig. 2). 
The weaker correlations in FeS, established by our observation via quantum oscillation measurements of minute enhancement in the effective mass over its non-interacting counterpart, both reduce the low-energy spin spectral weight and push up the energy scale for the $(1,1)$ excitations. 
The weaker correlations also imply that 
FeS is much closer to a BCS
superconductor,
which allow us to understand 
why the low-energy
spin excitations do not directly respond to superconductivity (Fig. 3).
These results for the isostructural and stoichiometric FeS highlight the strongly correlated nature of FeSe.
Indeed,
the electron spectral weight in FeSe mainly resides in the incoherent part, which induces quasi-local moments. 
The ensuing physics of frustrated magnetism not only yields the nematic order but also is manifested 
in the co-existing spin excitations at $(1,0)$ and $(1,1)$ wave vectors \cite{YuR,WangF}.
The strong correlations in FeSe also enhance the effective quasiparticle interactions in its superconducting state,
giving rise to a resonance spin excitation in FeSe \cite{Rahn,WangQ}.
As such, our findings elucidate both the origin of the nematic order and the nature of the superconductivity in FeSe.

\section{Methods}
Our quantum oscillation transport measurements on FeS were carried out at National High-magnetic Field laboratory in Tallahassee, Florida \cite{SI}.
Our inelastic neutron scattering measurements were carried out at the Fine-Resolution Fermi Chopper Spectrometer (SEQUOIA) at the Spallation Neutron Source, Oak Ridge National Laboratory and at the Multi Axis Crystal Spectrometer (MACS) at NIST Center for Neutron Research (NCNR), National Institute of Standards and Technology. Sample alignment for MACS and initial charactization is done at Spin Polarized Inelastic Neutron Spectrometer (SPINS), National Institute of Standards and Technology.
We have also performed neutron powder diffraction experiments on the BT-1, NCNR.
Single crystals of FeS (6.0 g for SEQUOIA and 6.5g for MACS) were grown using hydro-thermo method and characterizations of our samples are discussed in \cite{SI}. Pieces with size larger than 3*3 mm$^2$ were used in the neutron scattering experiment.
The elasto-resistance measurements were carried out using PPMS with a strain gauge attached on the piezo stack to measure strain at different temperatures. Measurements were performed by changing voltage on piezo stack and results presented here were scaled to actual strain in the sample.
To facilitate an easy comparison with the results on FeSe \cite{Rahn,WangQ,WangQa}, we used the orthorhombic notation with 
$a= b\approx 5.19$ \AA\ and $c=5.03$ \AA\ for FeS. In this notation, the stripe AF spin excitations for FeS occur at $(\pm 1,0,L)$ positions in reciprocal space. 
Samples are co-aligned in the $[H,0,L]$ scattering plane with a mosaic of 8$^\circ$. In the SEQUOIA experiment, the incident beam with $E_i=80,150$ meV was along the $c$-axis of the crystals. In the MACS experiment, 
$E_f=5$ meV was used for excitations above 1.6 meV and $E_f=3.7$ meV was used for excitations below 1.25 meV.  Details of DFT+DMFT calculations
are described in \cite{SI}.

\section{Acknowledgments}
The single crystal growth and neutron scattering work at Rice is supported by the
U.S. DOE, BES under contract no. DE-SC0012311 (P.D.). 
A part of the materials work at Rice is supported by the Robert A. Welch Foundation Grants No. C-1839 (P.D.).
The theoretical work at Rice is supported by the NSF Grant No.\ DMR-1611392 and the Robert A.\ Welch Foundation Grant No.\ C-1411 (Q.S.). Z.P.Y acknowledges financial support by the National Natural Science Foundation of China, Grant No. 11674030,  the National Key Research and Development Program of China under contract No. 2016YFA0302300.
 L.~B. is supported by DOE-BES through award DE-SC0002613.
The NHMFL is supported by NSF through NSF-DMR-1157490 and the
State of Florida. The use of ORNL's SNS was sponsored by the Scientific User Facilities Division, Office of BES, U.S. DOE.

\section{Author contributions}
Single crystal growth and neutron scattering experiments were carried out by H.M., J.G., R.Z. with 
assistance from M.F., M.S., Q.H., Y.S.,J.W.L., C.B. and P.D..
W.Y.W. and Y.S. performed ICP measurement.
Quantum oscillation measurements and analysis were performed by R.U.S. and L.B..
Theoretical understandings were performed by 
Z.P.Y. (DFT+DMFT), D.J.S, T.H., I.E. (DFT), and Q.S. (Electron correlations and magnetic frustration).
P.D. oversees the entire project.  The paper was written by
P.D., H.M., Z.P.Y., L.B., I.E., Q.S., and all authors made comments. 

\section{Competing interest}
The authors declare no competing financial interests.

\section{Additional information}
Correspondence and
requests for materials should be addressed to 
Z.P.Y (yinzhiping@bnu.edu.cn), L.B. (balicas@magnet.fsu.edu), or 
P.D. (e-mail: pdai@rice.edu)

\section{Figure Legends}

\begin{figure}[t]
\includegraphics[scale=1.3]{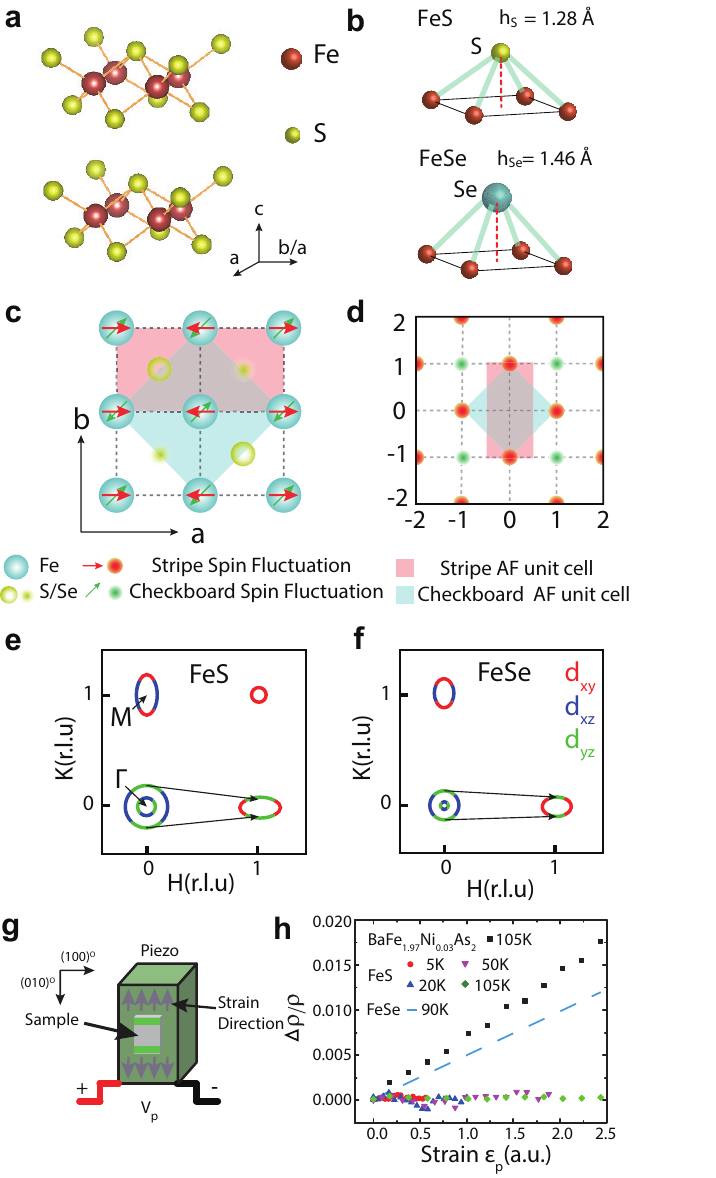}
\caption{{\bf Crystal structures, real/reciprocal spaces, Fermi surfaces, and transport measurements of FeS and FeSe.}
(a) The crystal structures of FeS or FeSe in orthorhombic notation.  The sulfur (S) can be fully substituted by selenium (Se) to form FeSe. 
(b) Schematic illustration of sulfur and selenium atoms in FeS and FeSe compounds. (c) Illustration of stripe (red) and checkerboard (green) 
static long range AF order in real space. The orthorhombic long-axis direction is along the $a$-axis for stripe AF order. (d) The corresponding positions for stripe and checkerboard orders and excitations in reciprocal space.  The areas of the Brillouin zones are marked as pink and blue, respectively.
Schematics of Fermi surfaces corresponding to FeS (e) and FeSe (f) 
with possible nesting wave vectors marked by arrows. The orbital components ({$d_{xz}$, $d_{yz}$, $d_{xy}$}) for different Fermi surfaces are 
shown in different colors. (g) Schematics of the setup used to measure elasto-resistance using a physical property measurement system \cite{Kuo2016}. 
(h) Strain dependence of the resistivity anisotropy $\Delta\rho/\rho= 2(\rho_a-\rho_b)/(\rho_a+\rho_b)$ for FeS, FeSe, and BaFe$_{1.97}$Ni$_{0.03}$As$_2$ at different temperatures.
}
\end{figure}

\begin{figure}[t]
\includegraphics[scale=.6]{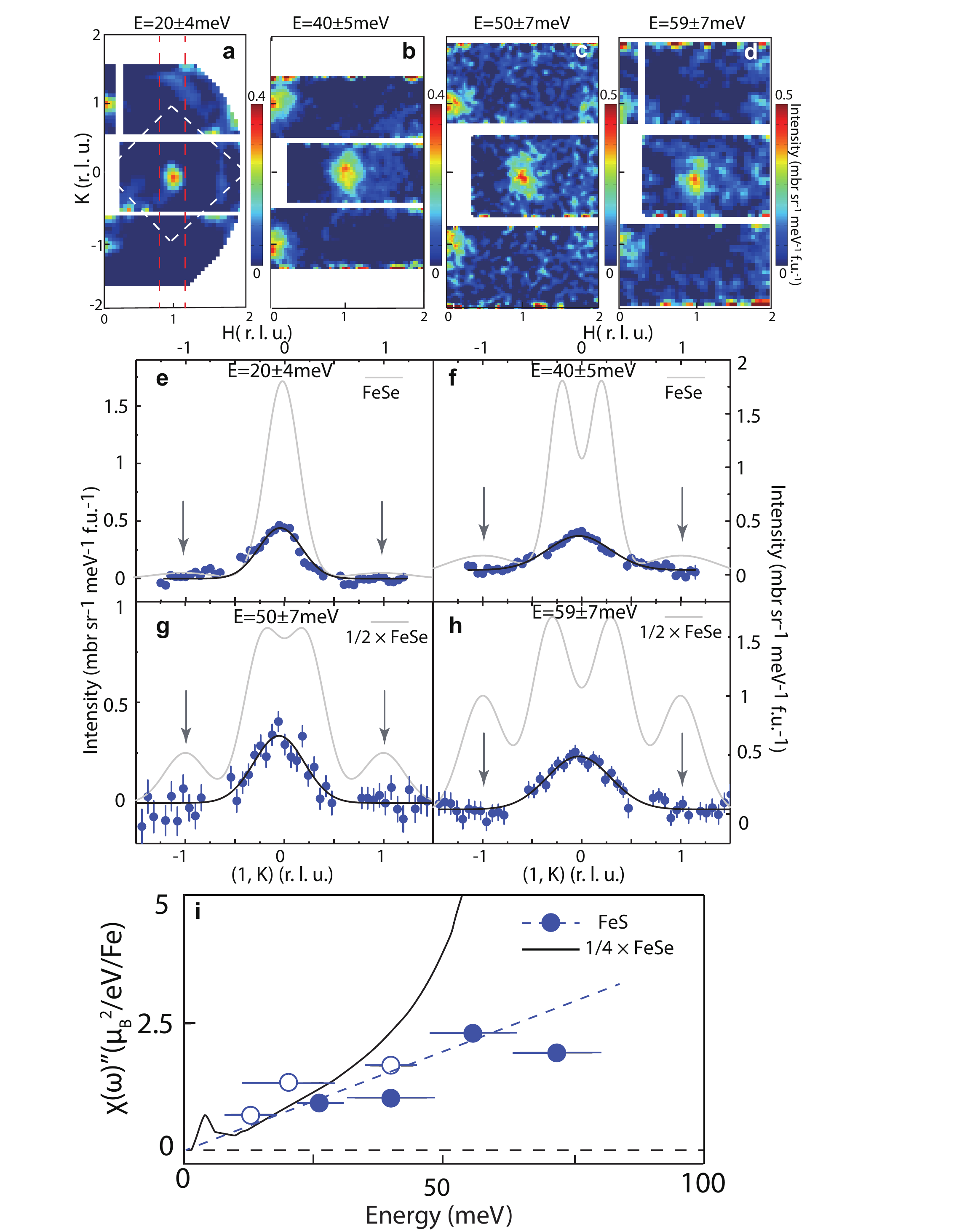}
\caption{ {\bf Spin excitations of FeS obtained by time-of-flight neutron spectroscopy.}
(a)-(d) Constant energy cuts measured at $T=4$ K at the energy transfers indicated on top of each panel. Red dashed lines in (a) indicate integrating area in reciprocal space for the 1D cuts in panels (e)-(h). 
The white dashed box indicates the area of integration to estimate the local dynamic susceptibility $\chi^{\prime\prime}(E)$ in panel (i). 
(e)-(h) Constant energy cuts through reciprocal space stripe AF wave vectors along the $[1,K,]$ direction at energies corresponding to panels (a)-(d).
The gray solid lines indicate fits to the  data extracted from excitations in FeSe at the same energy range \cite{WangQa}. 
Gray arrow indicates the checkerboard wave vector observed in FeSe, which is absent in FeS. (i) Comparison of the energy dependence of the  
local dynamic susceptibility $\chi^{\prime\prime}(E)$ for FeS and FeSe \cite{WangQa}.  The open and filled circles are data taken at $L=0.5, 1.5, \cdots$, and $0,1,2,\cdots$, respectively.  
 }
\end{figure}

\begin{figure}[t]
\includegraphics[scale=1.0]{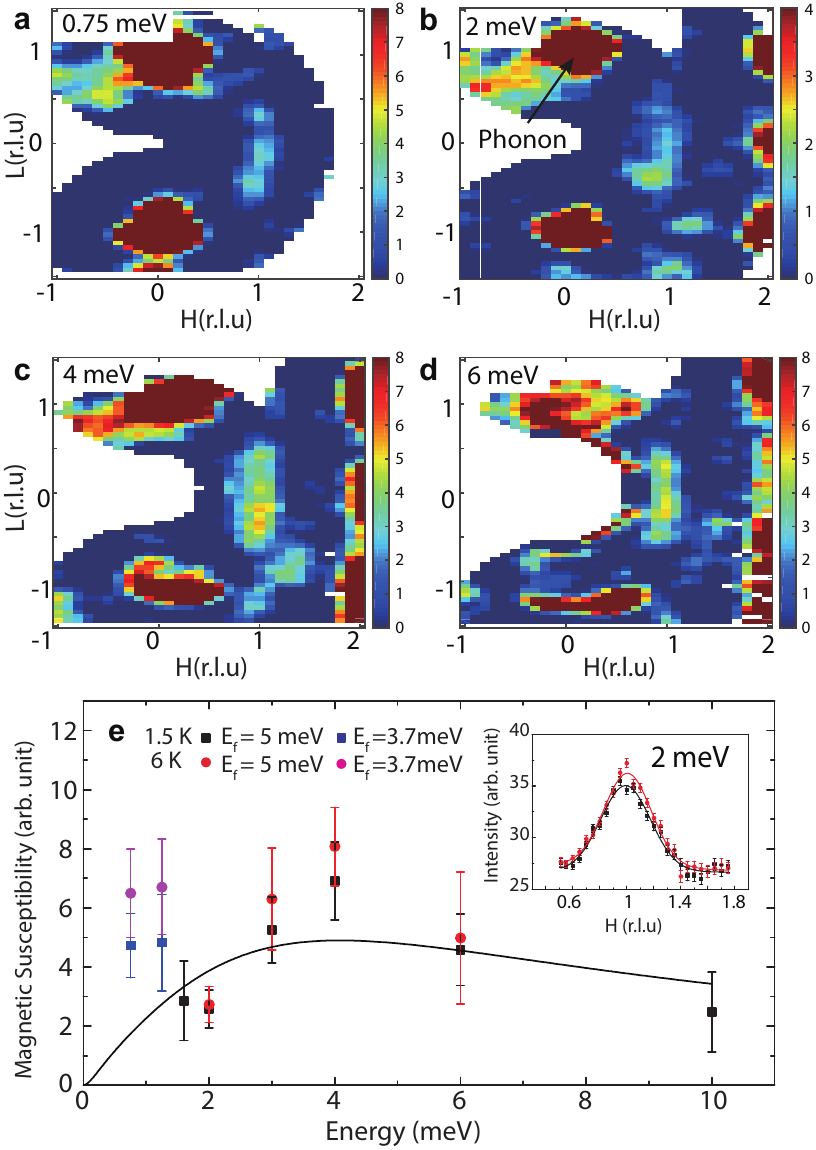}
\caption{ {\bf Temperature dependence of the low-energy spin excitations of FeS.} 
2D images of neutron scattering intensity in the $[H,0,L]$ scattering plane at energies of (a) $E = 1.25$, (b) 2, (c) 4, and (d) 6 meV \cite{SI}.
The high scattering intensity near the Bragg peak positions of $(0,0,\pm 1)$ is due to acoustic phonon scattering.  Spin excitations in FeS form a ridge of scattering 
centered at $(1,0,L)$ positions. (e) Temperature dependence of the stripe AF spin excitations at different energies below and above 
$T_c=4$ K, respectively.  Spin excitations are obtained by integrating $L$ from $-0.7\leq L\leq 0.7$, and
fitted with a linear background and a Gaussian peak as shown in the inset. The black line is a fit of the energy dependence of the spin excitations with a relaxation form 
$\chi^{\prime\prime}(E)=A\Gamma E/[(\Gamma/2)^2+E^2]$, where $\Gamma=8.2\pm 2.8$ meV.  Inset: $H$-scans at $E=2$ meV and at 1.5 K and 6 K respectively.
The solid lines are Gaussian fits on a linear background.
}
\end{figure}

\begin{figure}[t]
\includegraphics[scale=.6]{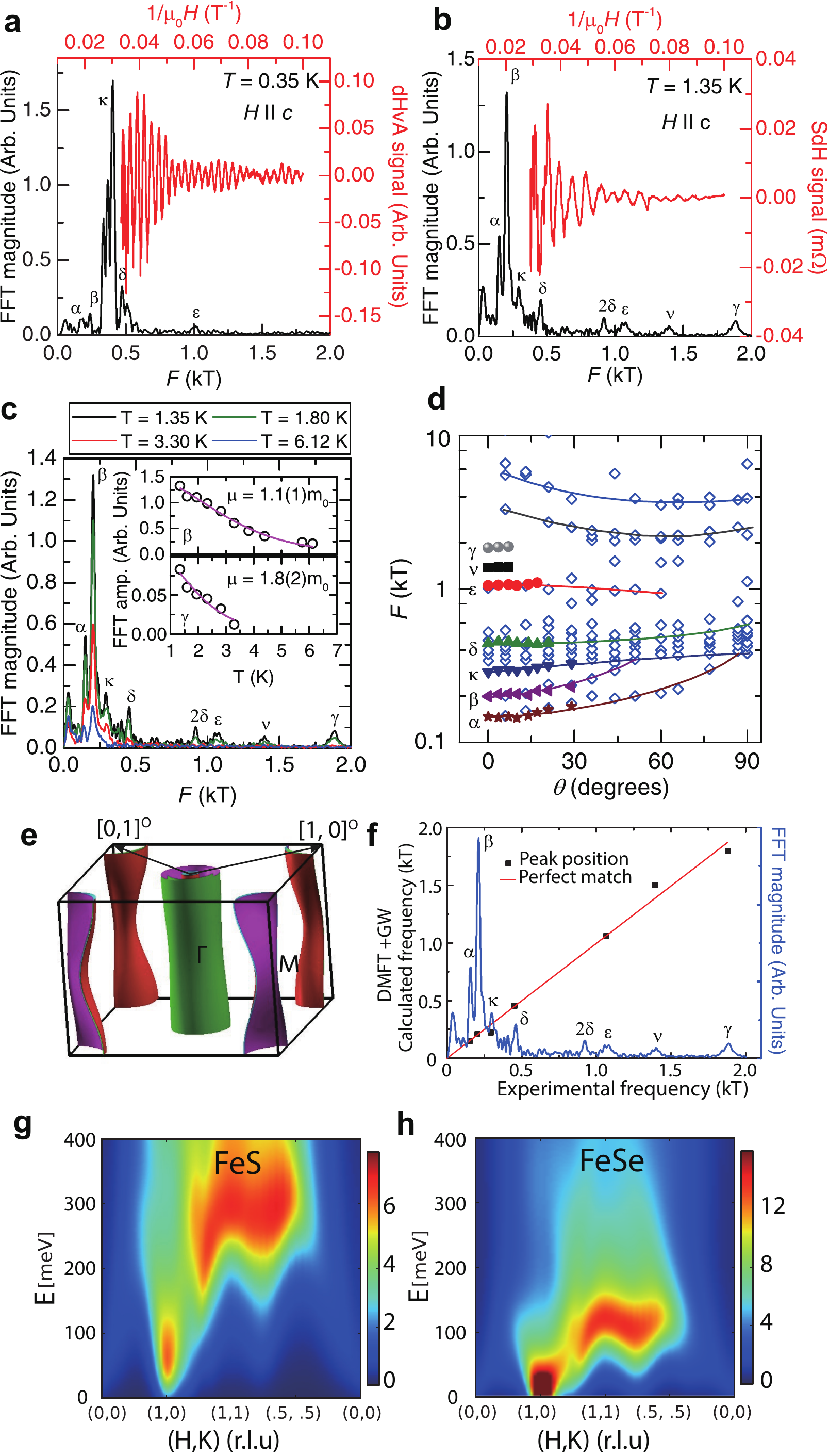}
\caption{ {\bf Quantum oscillations, Fermi surfaces and spin fluctuation spectra for FeS.}
(a) and (b) de Haas-van Alphen and Shubnikov-de Haas oscillations after background subtraction (red lines) with their associated Fast Fourier transformations (black lines) for magnetic fields applied parallel to the crystallographic $c$-axis. The dHvA signal was obtained at $T = 0.35\,\mathrm{K}$ and the SdH signal at $T= 1.35\,\mathrm{K}$, respectively. Greek letters ($\alpha$, $\beta$, $\gamma$, ...) indicate the most prominent peaks in the FFT spectra which can be assigned to extremal cross sectional areas of the Fermi surface. (c) Fourier transform spectra of the SdH oscillations for $H\parallel c$ at selected temperatures ranging from $1.35\,\mathrm{K}$ to $6.1\,\mathrm{K}$. Insets:  temperature dependence of the FFT amplitude for the $\beta$ and $\gamma$ orbits as well as their effective masses as obtained from the Lifshitz-Kosevich formalism (magenta lines). (d) Angular dependence of the dHvA (open diamonds) and SdH (filled symbols) frequencies. The dHvA and SdH measurements cover angles ranging from $H\parallel c$ ($\theta = 0^{\circ}$) to $H\perp c$ ($\theta = 90^{\circ}$). Solid lines represent a suggestion as to how the individual frequencies might evolve as a function of $\theta$. 
(e) DFT+DMFT calculated 3D Fermi surfaces \cite{SI}. The Fermi surface drawing is using the tetragonal structure and the corresponding orthorhombic directions $[1,0]$ and $[0,1]$ are marked by arrows.
(f) Comparison of frequencies of quantum oscillations with DFT+DMFT calculations.  Peak position (black square) is obtained by 
FFT the magnitude of resistance data shown in blue. 
The expected ground state spin excitations of FeS (g) and FeSe (h) as calculated by combined DFT+DMFT.
}
 \end{figure}

\end{document}